\documentclass[10pt,letterpaper]{article}
\usepackage{opex3}
\usepackage{color}
\usepackage{cite}

\begin{document}

\title{T-shaped Single-photon Router}

\author{Jing Lu$^{1}$, Z. H. Wang$^{2}$ and Lan Zhou$^{1,*}$}

\address{$^1$Synergetic Innovation Center of Quantum Effects and Applications,
Department of Physics, Hunan Normal University, Changsha 410081,
China\\$^2$ Beijing Computational Science Research Center, Beijing 100084, China}

\email{$^*$zhoulan@hunnu.edu.cn} 



\begin{abstract}
We study the transport properties of a single photon scattered by a two-level system (TLS)
in a T-shaped waveguide, which is made of two coupled-resonator waveguides (CRWs)--- an infinite
CRW and a semi-infinite CRW. The spontaneous emission of the TLS directs single photons from
one CRW to the other. Although the transfer rate is different for the wave incident from different
CRWs, due to the boundary breaking the translational symmetry, the boundary can enhance the transfer rate found in Phys. Rev. Lett. 111, 103604 (2013) and Phys. Rev. A 89, 013805 (2014), as the transfer rate
could be unity for the wave incident from the semi-infinite CRW.
\end{abstract}

\ocis{(270.0270) Quantum optics; (230.4555) Coupled resonators; (270.5585) Quantum information and processing.}



\section{Introduction}

Quantum channels and nodes are the building blocks of quantum networks~\cite%
{nwork}. Photons are natural carriers in quantum channels due to their
robustness in preserving quantum information during propagation. Quantum
channels are made of waveguides, which means that controlling
photons coherently in a confined geometry is of both fundamental and
practical importance for building quantum networks. Due to the negligible interaction
among individual photons, quantum devices at single-photon level have been
proposed based on the interaction of confined propagating fields with single atoms~\cite%
{FanQdev0,FanQdev1,FanQdev2,zhouQdev1,zhouQdev2,zhouQdev3,Supcavity1,Supcavity2,
Supcavity3,Supcavity4,cklaw09a,cklaw09b,3levNJP,attenuator,Diode,InMZ,Freconver1,Freconver2}.
As a quantum network has more than one quantum channel, a
multichannel quantum router~\cite{qrcyclic,qrlambda} for single photons has
been explored to transfer single photons from one quantum channel to the
other.

There are two ways to change the transport properties of particles in quantum channels:
the incorporation of impurities, or slight structural variations. Quantum routers
in~\cite{qrcyclic,qrlambda} have made good use of the arrangement of the energy configuration
of single atoms. In this paper, we propose a single-photon routing scheme using a
two-level system (TLS). Different from the studies in~\cite{qrcyclic,qrlambda}, where two infinite
one-dimensional (1D) coupled-resonator waveguides (CRWs) form a X-shaped waveguide, we consider
a slight structural variation of the two 1D CRWs, i.e., one 1D CRW is infinite and the other is
semi-infinite, which form a T-shaped waveguide~\cite{GWannierPC,TPRB03,TPRB05,TPRB11}. The systems
studies here, could be implemented using, e.g., artificial atoms~\cite{Nori01,Nori02,Nori03}
coupled to superconducting circuits~\cite{Nori04,Nori05,Nori06}. Aiming to answer the question
whether the boundary can enhance the transfer rate found in~\cite{qrcyclic,qrlambda}, we studied
the single-photon scattering process by the TLS, and found that: the spontaneous emission of the
TLS routes single photon from one CRW to the other; the probability for finding single photon in
one CRW is different for waves incident from different CRWs, due to the boundary breaking the
translational symmetry; The probability for finding single photon in the infinite CRW
could reach one for waves incident from the semi-infinite CRW, however, $50\%$ is the maximum transfer
rate in~\cite{qrcyclic,qrlambda}.

This paper is organized as follows: In Sec.~\ref{Sec:2}, the T-shaped waveguide with a TLS embedded
in its junction is introduced. In Sec.~\ref{Sec:3}, the single-photon scattering process is studied
for waves incident from different CRWs. Finally, we conclude with a brief summary of the results.


\section{\label{Sec:2}Model setup}

\begin{figure}[tbp]
\centering
\includegraphics[bb=29bp 528bp 460bp 764bp,width=8cm]{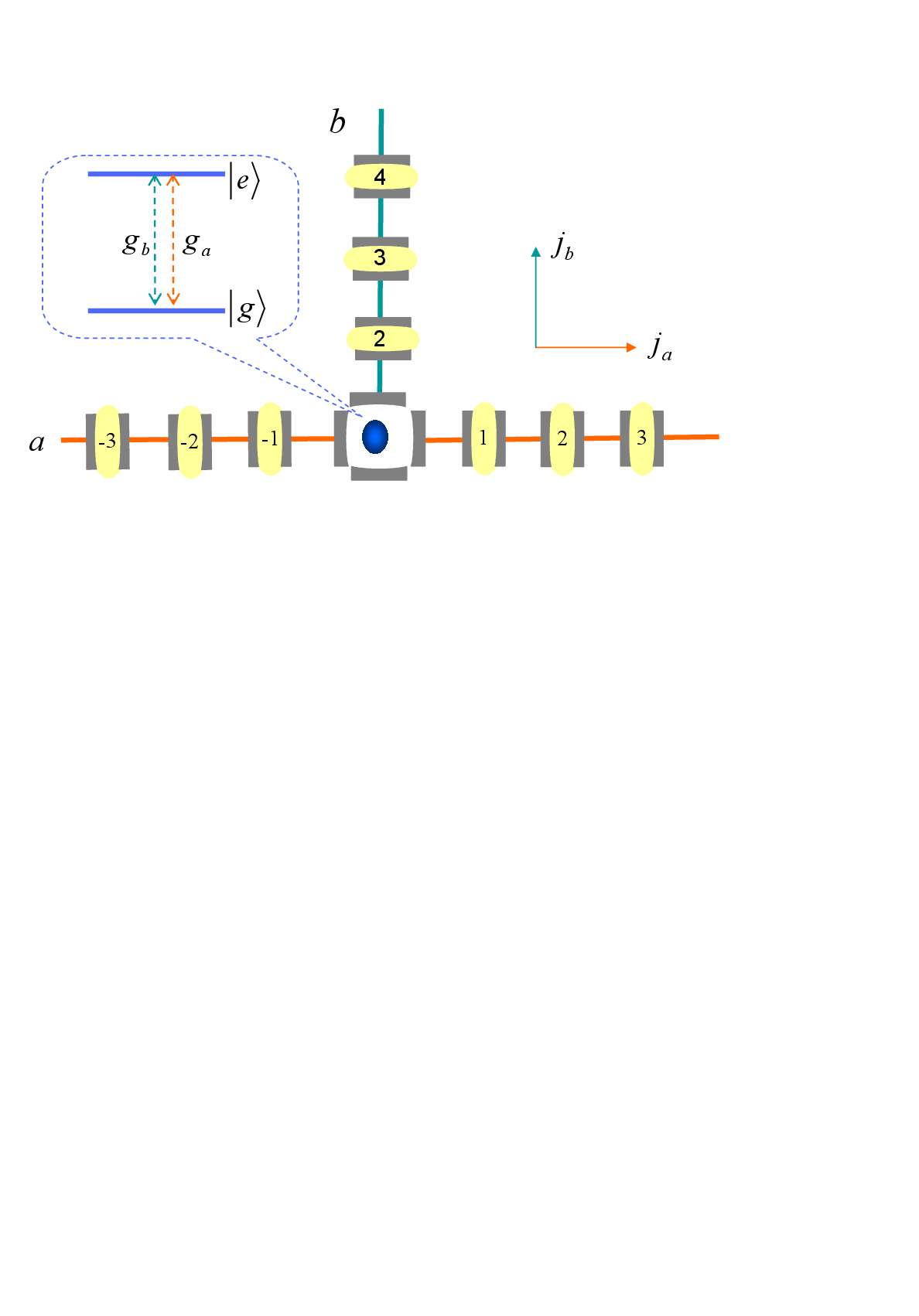}
\caption{Schematic view of quantum routing of single photons
in two channels made of one infinite and one semi-infinite CRWs. The
two-level atom characterized by $\left\vert g\right\rangle $, $\left\vert
e\right\rangle $ is placed at the cross point $j_{a}=0$ and $j_{b}=1$. CRW-$%
a $ (-$b$) couples to the atom through the transition $\left\vert
g\right\rangle \leftrightarrow \left\vert e\right\rangle $ with strength $%
g_{a}$ ($g_{b}$). There is no interaction among the $j_a=0$ and $j_b=1$ cavities}
\label{fig:1}
\end{figure}

A 1D CRW is made of single-mode cavities which are coupled to each other through the evanescent
tails of adjacent fields resulting in photon hopping among neighbouring cavities. Here, we consider
two 1D CRWs which form a T-shaped waveguide. As sketched in Fig.~\ref{fig:1}, coupled resonators
on the red (green) line construct the infinite (semi-infinite) CRW, which is called CRW-$a$ (CRW-$b$)
hereafter. We note that we have introduced a boundary in one of the CRWs in~\cite{qrcyclic,qrlambda}.
The purpose of classifying cavities in Fig.~\ref{fig:1} into the CRW-$a$ and -$b$ is
for the convenience to compare the results with those in~\cite{qrcyclic,qrlambda}.
The cavity modes of the two 1D CRWs are described by the annihilation operators $%
a_{j_{a}}$ and $b_{j_{b}}$, respectively. Here, subscripts $j_{a}=-\infty,\cdots ,+\infty $
and $j_{b}=1,\cdots ,+\infty $. The Hamiltonian of each CRW is described by a typical
tight-binding bosonic model. The free propagation of the photons in the T-shaped waveguide
is described by
\begin{equation}
H_{C}=\sum_{d=a,b}\sum_{j_{d}}\left[ \omega _{d}d_{j_{d}}^{\dagger
}d_{j_{d}}-\xi _{d}\left( d_{j_{d}}^{\dagger }d_{j_{d}+1}+\mathrm{h.c.}%
\right) \right] ,  \label{model-01}
\end{equation}%
where we have assumed that all cavities in the CRW $a$ $(b)$
have the same frequency $\omega _{a}$ $(\omega _{b})$ and the hopping
energies $\xi _{a}$ $(\xi _{b})$ between any two nearest-neighbor cavities
in the CRW-$a$ $(b)$ are the same. We note that there is no interaction between
the $j_a=0$ and $j_b=1$ cavities.

A TLS (an atom, a quantum dot, or a superconducting qubit) described by the
free Hamiltonian
\begin{equation}
H_{A}=\omega _{A}\left\vert e\right\rangle \left\langle e\right\vert
\label{model-02}
\end{equation}%
is placed at the node of the T junction, where $\omega _{A}$ is its energy splitting.
The TLS characterized by a ground state $\left\vert g\right\rangle $ and an excited state
$\left\vert e\right\rangle $ is coupled at a rate $g_{a}$ ($g_{b}$) to the cavity at
 $j_{a}=0$ ($j_{b}=1$). The interaction between the TLS and quantized electromagnetic modes reads
\begin{equation}
H_{AC}=g_{a}\sigma _{+}a_{0}+g_{b}\sigma _{+}b_{1}+h.c.  \label{model-03}
\end{equation}%
where the operators $\sigma_{\pm}$ are the ladder operators of the TLS. The dynamics of
the total system is governed by Hamiltonian $H=H_{C}+H_{A}+H_{AC}$.

The Hamiltonian~(\ref{model-01}) can be exactly diagonalized  as
$H_{C}=\sum_{d=a,b}E_{k_{d}}d_{k_{d}}^{\dagger }d_{k_{d}}$ by the Fourier transformations
$a_{k_{a}}=\frac{1}{\sqrt{2\pi }}\int dj_{a}a_{j_{a}}e^{ik_{a}j_{a}}$ and
$b_{k_{b}}=\sqrt{\frac{2}{\pi }}\int dj_{b}b_{j_{b}}\sin \left( k_{b}j_{b}\right) $ for
the CRW-$a$ and -$b$ respectively. The dispersion relation of both CRWs%
\begin{equation}
E_{k_{d}}=\omega _{d}-2\xi _{d}\cos k_{d}  \label{model-04}
\end{equation}%
is a cosine function of the wavenumber $k_{d}$ ($d=a,b$), which indicates
that each CRW possesses an energy band with bandwidth $4\xi _{d}$.
Consequently, two quantum channels (i.e., two broad continua of propagating
modes) are formed.


\section{\label{Sec:3}Single-photon quantum router}



Since the operator $N=\sum_{d=a,b}\sum_{j_{d}}d_{j_{d}}^{\dagger
}d_{j_{d}}+\sigma _{+}\sigma _{-}$ commutes with Hamiltonian $H$, the total
number of excitations is a conserved quantity. To find the scattering
equation in the single-excitation subspace, the eigenstate of the
full Hamiltonian is assumed to be
\begin{equation}
\left\vert E\right\rangle =\sum_{d=a,b}\sum_{j_{d}=-\infty }^{+\infty
}U_{j_{d}}^{\left[ d\right] }d_{j_{d}}^{\dag }\left\vert g0\right\rangle
+U_{e}\left\vert e0\right\rangle ,  \label{II-01}
\end{equation}%
where $\left\vert 0\right\rangle $ is the vacuum state of the T-shaped
waveguide, $U_{j_{a}}^{\left[ a\right] }$ ($U_{j_{b}}^{\left[ b\right] }$)
is the probability amplitudes of the photon in the $j_{a}$th ($%
j_{b}$th) cavity of the CRW-$a$ (CRW-$b$), and $U_{e}$ is the atomic
excitation amplitude. The eigenequation gives rise to a series of coupled
stationary equations for all amplitudes
\begin{eqnarray}
EU_{j}^{\left[ a\right] }& =&\omega _{a}U_{j}^{\left[ a\right] }-\xi
_{a}\left( U_{j-1}^{\left[ a\right] }+U_{j+1}^{\left[ a\right] }\right)
+g_{a}U_{e}\delta _{j0} \label{II-02a}\\
EU_{j}^{\left[ b\right] }& =&\omega _{b}U_{j}^{\left[ b\right] }-\xi
_{b}\left( U_{j-1}^{\left[ b\right] }+U_{j+1}^{\left[ b\right] }\right)
+\delta _{j1}g_{b}U_{e} \label{II-02b}\\
EU_{e}& =&\omega _{A}U_{e}+g_{a}U_{0}^{\left[ a\right] }+g_{b}U_{d}^{\left[ b%
\right] } \label{II-02c}
\end{eqnarray}
where $\delta _{mn}=1$ $(0)$ for $m=n$ $(m\neq n)$. Removing the atomic
amplitude leads to the scattering equations for single photons
\begin{eqnarray}
\left( E-\omega _{a}\right) U_{j}^{\left[ a\right] }& =&-\xi _{a}\left(
U_{j-1}^{\left[ a\right] }+U_{j+1}^{\left[ a\right] }\right)
 +\delta _{j0}\left[ V_{a}\left( E\right) U_{0}^{\left[ a\right] }+G\left(
E\right) U_{1}^{\left[ b\right] }\right] \label{II-03a} \\
\left( E-\omega _{b}\right) U_{1}^{\left[ b\right] }& =&-\xi _{b}U_{2}^{\left[
b\right] }+G\left( E\right) U_{0}^{\left[ a\right] }+V_{b}\left( E\right)
U_{1}^{\left[ b\right] } \label{II-03b}\\
\left( E-\omega _{b}\right) U_{j}^{\left[ b\right] }& =&-\xi _{b}\left(
U_{j-1}^{\left[ b\right] }+U_{j+1}^{\left[ b\right] }\right) (j\geq 2).
\label{II-03c}
\end{eqnarray}
The coupling between the TLS and CRWs gives rise to the energy-dependent
deltalike potentials with strength $V_{d}\left( E\right) =g_{d}^{2}/\left(
E-\omega _{A}\right) $ and the effective dispersive coupling strength $%
G\left( E\right) =g_{a}g_{b}/\left( E-\omega _{A}\right) $ between two CRWs,
which are highly localized.

In this paper, we are interested in routing single photons from one CRW to
the other, i.e., the TLS acts as a multichannel quantum router. According
to the results in~\cite{qrcyclic,qrlambda}, we set $\omega _{a}=\omega _{b}=\omega $
and $\xi_{a}=\xi _{b}=\xi $ in the following discussion. Since the boundary
of the CRW-$b$ breaks the translational symmetry, we solve Eqs.(\ref{II-03a}-\ref{II-03c})
for waves incoming from both CRWs separately.

\subsection{Single photons incident from the infinite CRW-$a$}

For a photon with wavenumber $k$ incident along the $j_{a}$
axis onto the T-shaped waveguide, it will be absorbed by the TLS, which
transits from its ground state to its excited state. Since the excited state
is coupled to the continua of states, the excited TLS will emit a photon
spontaneously into the propagating state of either CRW-$a$ or CRW-$b$. Then
a scattering process of single photons is completed, i.e., waves encoutering
the TLS result in reflected, transmitted, and transferred waves with the
same energy $E=\omega -2\xi \cos k$. The boundary forces the photon within
a certain region of space. We search for a solution of Eqs.(\ref{II-03a})-(\ref{II-03c})
in the form
\begin{eqnarray}
U_{j_{a}}^{\left[ a\right] }& =\left\{
\begin{array}{c}
e^{ikj_{a}}+re^{-ikj_{a}},\quad j_a<0 \\
te^{ikj_{a}},\quad j_a>0%
\end{array}%
\right. \label{II-04a} \\
U_{j_{b}}^{\left[ b\right] }& =\left\{
\begin{array}{c}
t^{b}e^{ikj_{b}},\quad j_{b}>1 \\
A\sin kj_{b},\quad j_{b}=1%
\end{array}%
\right. \label{II-04b}
\end{eqnarray}
where $r$, $t$, and $t^{b}$ are the reflection, transmission, and transfer
amplitudes respectively, and $A$ is the amplitude at the boundary cavity $%
j_{b}=1$. Substitution of Eqs.(\ref{II-04a})-(\ref{II-04b}) into Eqs.(\ref{II-03a})-(\ref{II-03c}), after
some algebra, we obtain the relation $U_{0}^{\left[ a\right] }=t=1+r$, and
\begin{eqnarray}
t& =\frac{v_g\left( E-\omega _{A}\right) +\sin \left( 2k\right)
g_{b}^{2}+i2g_{b}^{2}\sin ^{2}k}{v_g\left( E-\omega _{A}\right)
+g_{b}^{2}\sin \left( 2k\right) +i\left( g_{a}^{2}+2g_{b}^{2}\sin
^{2}k\right) }, \label{II-05a}\\
t^{b}& =\frac{-2g_{a}g_{b}\sin k}{v_g\left( E-\omega _{A}\right)
+g_{b}^{2}\sin \left( 2k\right) +i\left( g_{a}^{2}+2g_{b}^{2}\sin
^{2}k\right) },\label{II-05b}
\end{eqnarray}
where the group velocity $v_g=2\xi \sin k$.
It can be verified that the scattering amplitudes satisfy $\left\vert
t\right\vert ^{2}+\left\vert r\right\vert ^{2}+\left\vert t^{b}\right\vert
^{2}=1$, which indicates probability conservation for the photon. It can be
found in Eqs.~(\ref{II-05a})-(\ref{II-05b}) that when $g_{b}=0$, $t^{b}=0$, and the
transmission amplitude $t$ is the same as the one obtained in~\cite%
{zhouQdev1}, where the system showed the resonant scattering at energy $%
E=\omega _{A}$ and $\Gamma_a(E)=g_{a}^{2}/v_g$ is regarded as the width of
the resonance, or the decay rate of the TLS into the modes of the CRW-$a$.
However, there is no frequency shift of the atom. It can be found in Eq.(\ref{II-05b})
that the coupling between the waveguide and the TLS that plays the important
role on transferring single photon from one quantum channel to the other.

\begin{figure}[tbp]
\centering
\includegraphics[width=9cm]{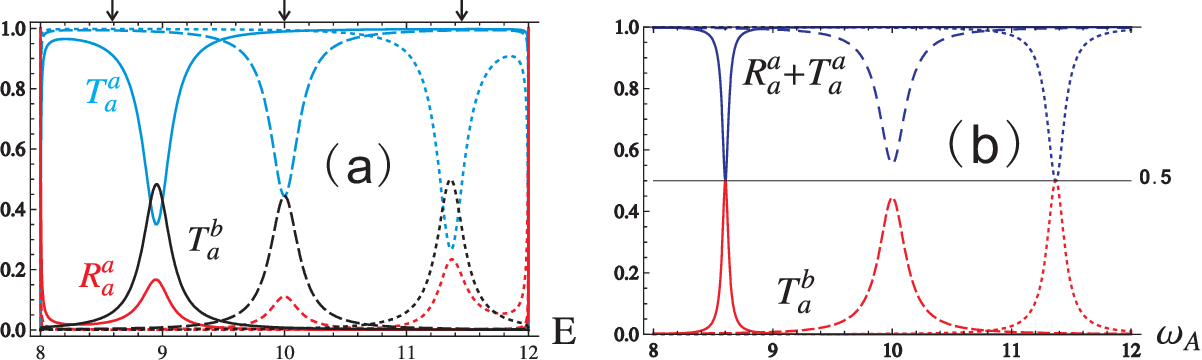}
\caption{The probability for finding single photons. (a) The transmission $T_a^a(E)$ (blue line),
reflection $R_a^a$ (red line) $T^{b}_a(E)$ (black line) as a function of the incident energy $E$,
where the parameters are set as follow: $\omega _{A}=9$ for solid line, $\omega _{A}=10$ for dashed
line, $\protect\omega _{A}=11.3$ for dotted line, and the coupling strength $g_{a}=g_{b}=0.3$,
(b) The coefficient $R_a^a(E)+T_a^a(E)$ (blue line), $T^{b}_a(E)$ (red line) as a function of the atomic energy splitting
$\omega _{A}$, where the parameters are set as follow: $g_{a}=g_{b}=0.15,E=10-\sqrt{2}$ for solid line,
$g_{a}=g_{b}=0.3,E=10$ for dotted line, $g_{a}=g_{b}=0.25,E=10+\sqrt{2}$ for dashed line. All the
parameters are in units of $\xi $, and we always set $\omega =10$. The arrows in (a) indicate
the position of the incident energy E of the solid, dashed and dotted lines in (b).}
\label{fig:2}
\end{figure}
In Fig.~\ref{fig:2}, we plot transmission $T_a^a(E)=\left\vert t\right\vert ^{2}$ (blue line), reflection
$R_a^a=\left\vert r\right\vert ^{2}$ (red line), $T^{b}_a(E)=\left\vert t^{b}\right\vert ^{2}$ (black line)
as well as coefficients $T_{a}^{a}+R_{a}^{a}$ (blue line) and $T_{a}^{b}$ (red line) as a function of the
incident energy $E$. We note that $T_{a}^{a}+R_{a}^{a}$ ($T_{a}^{b}$) gives the probability for finding a single
photon in the CRW-$a (b)$. It can be observed that: 1) the maximum transfer rate is $50\%$; 2) Although the
magnitudes of the probabilities are dependent on the atomic energy splitting, the energy splitting of the TLS
is no longer the position of the peak; 3) The product of the coupling strengths determines the width of the
lineshape. Comparing to the observations in~\cite{qrcyclic,qrlambda}, the only difference is the resonant-scattering
energy, i.e., the second observation here.

\subsection{Single photons incident from the semi-infinite CRW-$b$}

Now, we consider that a plane Bloch wave of a single photon is launched from the
upper to the bottom along the $-j_{b}$ axis into the semi-infinite CRW-$b$
with the dispersion relation $E=\omega -2\xi \cos k$. When the traveling
photon arrives at the node of the T junction, it is either absorbed by the
TLS or reflected by the boundary. The fraction reflected by the boundary
propagates along the positive $j_{b}$ axis. The portion absorbed by the
TLS is reemitted into the waveguide. The emitted radiation propagates into two
directions in the CRW-$a$, namely the forward and backward directions along the
CRW-$a$. In the CRW-$b$, the TLS radiates the photon to the upwards and downwards,
the photon originally radiated to the downwards is retroreflected to the TLS, since
the termination of the CRW-$b$ imposes a hard-wall boundary condition on the field
which behaves as a perfect mirror. The probability amplitudes in the asymptotic
regions are given by
\begin{eqnarray}
U_{j_{a}}^{\left[ a\right] }& =& \left\{
\begin{array}{c}
t_{l}^{a}e^{-ikj_{a}},\quad j_{a}<0 \\
t_{r}^{a}e^{ikj_{a}},\quad j_{a}>0%
\end{array}%
\right. \label{II-06a} \\
U_{1}^{\left[ b\right] }& =& A\sin k,\quad j_{b}=1\label{II-06b} \\
U_{j_{b}}^{\left[ b\right] }& =& e^{-ikj_{b}}+r^{b}e^{ikj_{b}},\quad j_{b}>1,
\label{II-06c}
\end{eqnarray}
where $t_{r}^{a}$ ($t_{l}^{a}$) and $r^{b}$ have the meaning of the forward
(backward) transfer and reflected amplitudes in the TLS-free region. After
some algebra, we obtain $t_{r}^{a}=t_{l}^{a}\equiv t^{a}$, which guarantees
the no discontinuity in the value of the wave function.
With Eqs.~(\ref{II-03a}-\ref{II-03c}), the transfer and reflected amplitudes read
\begin{eqnarray}
t^{a}& =& \frac{-2g_{a}g_{b}\sin k}{v_g\left( E-\omega _{A}\right)
+g_{b}^{2}\sin \left( 2k\right) +i\left( 2g_{b}^{2}\sin
^{2}k+g_{a}^{2}\right) }, \label{II-07a}\\
r^{b}& =& -\frac{v_g\left( E-\omega _{A}+\frac{g_{b}^{2}}{\xi }%
e^{-ik}\right) +i g_{a}^{2}}{v_g\left( E-\omega _{A}+\frac{g_{b}^{2}}{%
\xi }e^{ik}\right)+i g_{a}^{2}}.
\label{II-07b}
\end{eqnarray}
In this case, the probability for finding the photon in CRW-$a$ ($b$)
is equal to the transfer rate $T_{b}^{a}\equiv 2\left\vert
t^{a}\right\vert ^{2}$ (reflectance $R_{b}^{b}\equiv \left\vert r^{b}\right\vert ^{2}$).
It is easily to find that $T_{b}^{a}+R_{b}^{b}=1$ which guarantees the
probability conservation for the incident photon.

When $g_{a}=0$, the transfer amplitudes vanish. According to the probability
conservation, all the incident waves get perfectly reflected. However, the phase
of the reflected amplitude could be nonzero. As $g_{a}=0$, the system becomes
a semi-infinite CRW with a TLS inside. The hard-wall boundary due to the CRW
termination reflects all the incident waves. The absorption and emission of
single photons by the TLS introduce the phase different from $\pi $ to the
reflected amplitude, which is originated from the radiative properties of the
TLS. It is well-known that spontaneous emission of a TLS depends on the electromagnetic
vacuum environment that the atom is subjected to. Here, the boundary modifies the
radiation field and acts back onto the TLS. According to the method of
images~\cite{PRE73(06)}, the radiated photon is reflected by the virtual $%
j_{b}=0$ cavity. Consequently, the excited state of the TLS is dressed by
its own radiation field with the Lamb shift $\Delta \left( E\right)=g_{b}^{2}\cos k/\xi $
and the decay rate $\Gamma_b \left( E\right)=2 g_{b}^{2}\sin^2 k/v_g $ of the TLS
into the waveguide modes of the CRW-$b$, where the group velocity $v_g=2\xi\sin k$. In
the weak-coupling limit $g_{b}\rightarrow 0^{+}$, the transition frequency
of the TLS is renormalized as $\omega _{A}+\Delta \left( \omega _{A}\right) $.
Actually, the changes in the radiative rates of the TLS can be qualitatively
understood by the following consideration. First, let us explain how the factor
two appears in $\Gamma_b$. In an infinite CRW, the TLS radiates amplitude $\alpha$
to the upward, so does it to the downward. Therefore the total rate of the atomic
energy loss into an infinite CRW is proportional to $2|\alpha|^2$. In the presence
of a perfect mirror, light originally radiated to the downward is reflected back
toward the TLS and interferes with the light originally radiated to the upward.
For constructive interference, the total amplitude is $2\alpha$. And the total rate
of the atomic energy loss into an semi-infinite CRW is proportional to $4|\alpha|^2$,
which is twice larger than the total rate of the single atom
embedded in the infinite CRW. Now, we concern the factor $\sin^2 k$ appearing in $\Gamma_b$. For a
TLS inside an infinite CRW, $g_b$ is the coupling strength between the continuum
and the TLS. Then, we obtain $2|\alpha|^2=g_b^2/v_g$. However, for the TLS inside a
semi-infinite CRW, the coupling strength between the continuum and the TLS is modified
as $g_b\sin k$ by the Fourier transformation. Consequently, we obtain $4|\alpha|^2=\Gamma_b$. 
\begin{figure}[tbp]
\centering
\includegraphics[width=10cm]{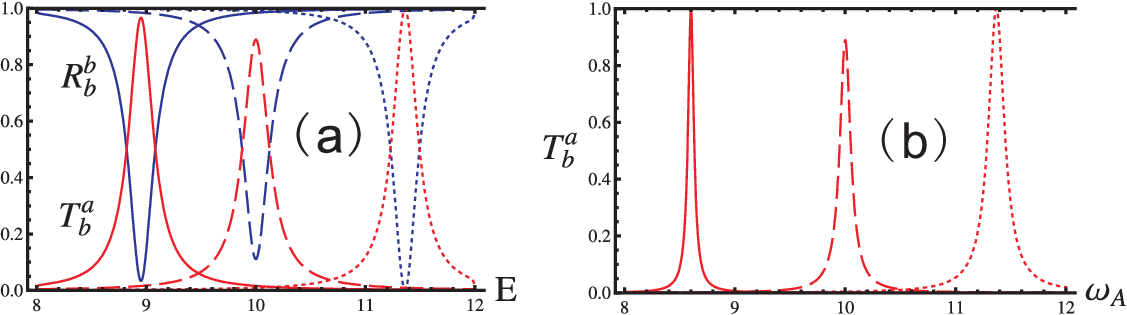}
\caption{The  probabilities for finding single photons in the
CRW-$a$ (blue line) and the CRW-$b$ (red line). (a) The probabilities as a
function of the incident energy $E$. The parameters are the same in Fig.~\ref{fig:2}(a).
(b) The coefficient $T^a_{b}(E)$ as a function of the atomic
energy splitting $\protect\omega _{A}$, The parameters are the same in Fig.\ref{fig:2}(b).}
\label{fig:3}
\end{figure}

With $g_{a}\neq 0$, the CRW-$a$ provides an extra channel for the radiated photon. Transferring
becomes possible after the incident photon is absorbed by the TLS. Therefore, the transfer
rate should be related to both the coupling strength $g_a$ and the modified coupling strength
$g_b\sin k$ by the boundary, which is why $T_b^a$ in Eq.~(\ref{II-07a}) has the product of
$g_a$ and $g_b\sin k$ in its numerator. The coupling of the TLS to the extra channel introduces
additional atomic energy loss, which is characterized by the decay rate $\Gamma_a$. Hence, the decay
rate of the TLS is the sum of $\Gamma_a$ and $\Gamma_b$, which construct the imaginary part of
the denominator in Eqs.~(\ref{II-05a})-(\ref{II-05b}) and (\ref{II-07a})-(\ref{II-07b}). Since the energy-level shift caused by
the CRW-$a$ is zero, the real part of the denominator in Eqs.~(\ref{II-05a})-(\ref{II-05b}) and (\ref{II-07a})-(\ref{II-07b})
only contains the atomic transition energy $\omega_A$ and the Lamb shift $\Delta(E)$ introduced
by the semi-infinite CRW. One can also observe that the transfer amplitudes in Eqs. (\ref{II-05b})
and (\ref{II-07a}) have the same expression. However, the transfer rate $T_{b}^{a}$ is twice larger
than $T_{a}^{b}$. According to studies in the previous section, the maximum $T_{b}^{a}$ could be one.
By comparison, we plot the probabilities for finding the photon in each CRW in Fig.\ref{fig:3} with
the parameters same to Fig.~\ref{fig:2}. Obviously, the reflectance $R_{b}^{b}$ could be lower than
$50\%$, even down to zero. Actually, it is not difficult to find that a peak of the transfer rate occurs
when the incident energy satisfies the resonant condition $E-\omega _{A}+\Delta \left( E\right)=0$
for the given coupling strengths. The results in~\cite{qrcyclic,qrlambda} told us that the
decay rates should be equal to further improve the transfer rate, i.e., the incident energy should
further satisfies $2g_{b}^{2}\sin^{2}k=g_{a}^{2}$ (called decay-match condition), which requires
that $g_{a}\leq \sqrt{2}g_{b}$. The above discussion told us that when the coupling and hopping
strengths are fixed, one can adjust $\omega_A$ to achieve the maximal value one of the transfer
rate for the photon with a given incident energy.


\section{Conclusion}

In this work, we have analytically studied the scattering process of single photons in a T-shaped
waveguide with a TLS embedded in the node of the T junction, where the T-shaped waveguide is made
of two CRWs ---an infinite CRW (refer to CRW-$a$) and a semi-infinite CRW (refer to
CRW-$b$). Comparing to the observations in~\cite{qrcyclic,qrlambda} with the X-shaped waveguide
which made of two noninteracting infinite CRWs, the boundary introduces new physical features: 1)
There are transmission and reflection in the incident CRW-$a$, but only reflection in the incident
CRW-$b$. Consequently, the probabilities for finding the photon in CRW-$a$ and CRW-$b$ have different
expressions, for example, $R_a^a+T_a^a$ in CRW-$a$ and $T_a^b$ in CRW-$b$ for waves incident from the
CRW-$a$, and $T_b^a$ in CRW-$a$ and $R_b^b$ in CRW-$b$ for waves incident from the CRW-$b$. 2)
The probability for successfully transferring the photon is different. The maximum magnitude of the
transfer rate is $50\%$ for waves incident from the CRW-$a$, however, $100\%$ for waves incident
from the CRW-$b$. Since $50\%$ is the maximum probability for transferring single photons from one CRW to
the other in the previous schemes~\cite{qrcyclic,qrlambda}, the boundary indeed enhance the transfer
probability. We note that since the linear response of the TLS is considered, the results could apply
 to the equivalent classical model of coupled oscillators.

\section*{Acknowledgments}
This work is supported by NSFC No.~11374095, No.~11422540, No.~11434011, No.~11575058; NBRPC No.~2012CB922103;
Hunan Provincial Natural Science Foundation of China (11JJ7001, 12JJ1002). Z. H. Wang is supported by Postdoctoral
Science Foundation of China (under Grant No. 2014M560879).


\begin{thebibliography}{99}
\bibitem{nwork} H.J. Kimble, ``The quantum internet,'' Nature (London) \textbf{453}, 1023-1030 (2008).

\bibitem{FanQdev0} J.T. Shen and S. Fan, ``Coherent single photon transport in a one-dimensional waveguide coupled
with superconducting quantum bits,'' Phys. Rev. Lett. \textbf{95}, 213001 (2005).
\bibitem{FanQdev1} J.T. Shen and S. Fan, ``Theory of single-photon transport in a single-mode waveguide. I.
coupling to a cavity containing a two-level atom,'' Phys. Rev. A \textbf{79}, 023837 (2009).
\bibitem{FanQdev2} J.T. Shen and S. Fan, ``Theory of single-photon transport in a single-mode waveguide. II.
coupling to a whisperinggallery resonator containing a two-level atom,'' Phys. Rev. A \textbf{79}, 023838 (2009).

\bibitem{zhouQdev1} L. Zhou, Z.R. Gong, Y.-x. Liu, C.P. Sun, and F. Nori, ``Controllable scattering of a single
photon inside a one-dimensional resonator waveguide,'' Phys. Rev. Lett. \textbf{101}, 100501 (2008).
\bibitem{zhouQdev2} L. Zhou, S. Yang, Yu-xi Liu, C.P. Sun and F. Nori, ``Quantum zeno switch for single-photon
coherent transport,'' Phys. Rev. A \textbf{80}, 062109 (2009).
\bibitem{zhouQdev3} Q. Li, L. Zhou, and C.P. Sun, ``Waveguide quantum electrodynamics: controllable channel
from quantum interference,'' Phys. Rev. A \textbf{89}, 063810 (2014).

\bibitem{Supcavity1} L. Zhou, H. Dong, Y.-x. Liu, C.P. Sun, and F. Nori, ``Quantum supercavity with atomic
mirrors,'' Phys. Rev. A \textbf{78}, 063827 (2008).
\bibitem{Supcavity2} Z.R. Gong, H. Ian, L. Zhou, and C.P. Sun, ``Controlling quasibound states
in a one-dimensional continuum through an electromagnetically-induced-transparency mechanism,''
Phys. Rev. A \textbf{78}, 053806 (2008).
\bibitem{Supcavity3} H. Dong, Z.R. Gong, H. Ian, L. Zhou and C.P. Sun, ``Intrinsic cavity QED and
emergent quasinormal modes for a single photon,'' Phys. Rev. A \textbf{79}, 063847 (2009).
\bibitem{Supcavity4} J.Q. Liao, Z.R. Gong, L. Zhou, Y.X. Liu, C.P. Sun and F. Nori, ``Controlling the transport
of single photons by tuning the frequency of either one or two cavities in an array of coupled cavities,''
Phys. Rev. A \textbf{81}, 042304 (2010).

\bibitem{cklaw09a} T.S. Tsoi and C. K. Law, ``Quantum interference effects of a single photon
interacting with an atomic chain inside a one-dimensional waveguide,'' Phys. Rev. A \textbf{78}, 063832
(2008).
\bibitem{cklaw09b} T.S. Tsoi and C. K. Law, ``Single-photon scattering on $\Lambda$-type three-level
atoms in a one-dimensional waveguide,'' Phys. Rev. A \textbf{80}, 033823 (2009).

\bibitem{3levNJP} D Witthaut and A.S. S{\o }ensen, ``Photon scattering by a three-level
emitter in a one-dimensional waveguide,'' New J. Phys. \textbf{12}, 043052 (2010).

\bibitem{attenuator} C.-H. Yan, L.-F. Wei, W.-Z. Jia, and J.-T. Shen, ``Controlling resonant
photonic transport along optical waveguides by two-level atoms,'' Phys. Rev. A \textbf{84}, 045801 (2011).

\bibitem{Diode} Y.-C. Shen, M. Bradford, and J.-T. Shen, ``Single-photon diode by exploiting the
photon polarization in a waveguide,'' Phys. Rev. Lett. \textbf{107}, 173902 (2011).

\bibitem{InMZ} L. Zhou, Y. Chang, H. Dong, L.M. Kuang, and C.P. Sun, ``Inherent Mach-Zehnder
interference with ¡°which-way¡± detection for single-particle scattering in one dimension,'' Phys.
Rev. A \textbf{85}, 013806 (2012).

\bibitem{Freconver1} M. Bradford, K.C. Obi, and J.-T. Shen, ``Efficient single-photon frequency
conversion using a sagnac interferometer,'' Phys. Rev. Lett. \textbf{108}, 103902 (2012).
\bibitem{Freconver2} Z.H. Wang, L. Zhou, Y. Li, and C. P. Sun, ``Controllable single-photon frequency
converter via a one-dimensional waveguide,'' Phys. Rev. A \textbf{89}, 053813 (2014).

\bibitem{qrcyclic} L. Zhou, L.P. Yang, Y. Li, and C.P. Sun, ``Quantum routing of single photons
with a cyclic three-level system,'' Phys. Rev. Lett. \textbf{111}, 103604 (2013).

\bibitem{qrlambda} J. Lu, L. Zhou, L.M. Kuang, and F. Nori, ``Single-photon router: Coherent
control of multichannel scattering for single photons with quantum interferences,'' Phys. Rev. A
\textbf{89}, 013805 (2014).

\bibitem{GWannierPC} J.P. Albert, C. Jouanin, D. Cassagne, and D. Bertho, ``Generalized Wannier
function method for photonic crystals,'' Phys. Rev. B \textbf{61}, 4381-4384 (2000).

\bibitem{TPRB03} N. Malkova and V. Gopalan, ``Strain-tunable optical valves at T-junction waveguides
in photonic crystals,'' Phys. Rev. B \textbf{68}, 245115 (2003).

\bibitem{TPRB05} T. Herrle, S. Schmult, M. Pindl, U.T. Schwarz, and W. Wegscheider, ``T-shaped waveguides
for quantum-wire intersubband lasers,'' Phys. Rev. B \textbf{72}, 035316 (2005).

\bibitem{TPRB11} E. Bulgakov and A. Sadreev, ``Symmetry breaking in a T-shaped photonic waveguide
coupled with two identical nonlinear cavities,'' Phys. Rev. B \textbf{84}, 155304 (2011).


\bibitem{Nori01} I. Buluta, S. Ashhab and F. Nori, ``Natural and artificial atoms for quantum computation,''
 Reports on Progress in Physics \textbf{74} 104401 (2011).
\bibitem{Nori02} I. Buluta, F. Nori, `` Quantum simulators,'' Science \textbf{326}, 108-111 (2009).
\bibitem{Nori03} I. M. Georgescu, S. Ashhab and  F. Nori, `` Quantum simulators,'' Rev. Mod. Phys. \textbf{86}, 153-185 (2014).
\bibitem{Nori04} J.Q. You and F. Nori `` Superconducting circuits and quantum information,'' Physics Today \textbf{58} (11), 42-47 (2005).
\bibitem{Nori05} J.Q. You and F. Nori `` Atomic physics and quantum optics using superconducting circuits,'' Nature \textbf{474}, 589-597 (2011).
\bibitem{Nori06} Z.L. Xiang, S. Ashhab, J.Q. You and F. Nori ``Hybrid quantum circuits: superconducting circuits interacting
with other quantum systems,'' Rev. Mod. Phys. \textbf{85}, 623-653 (2013).

\bibitem{PRE73(06)} K.G. Makris and D.N. Christodoulides, ``Method of images in optical discrete systems,''
Phys. Rev. E \textbf{73}, 036616 (2006).

\end{thebibliography}
\end{document}